\documentclass[%
 reprint,
 superscriptaddress,
 amsmath,amssymb,
 aps, pre
]{revtex4}

\usepackage[usenames,dvipsnames]{xcolor}
\usepackage{amsthm}
\usepackage{amssymb}

\usepackage{graphicx}
\usepackage{dcolumn}
\usepackage{bm}
\usepackage[T2A]{fontenc}
\usepackage[utf8]{inputenc}
\usepackage[russian,english]{babel}
\usepackage{comment}
\addto\captionsenglish{}
\usepackage{bigstrut}
\usepackage{float}

\usepackage[unicode, colorlinks, urlcolor=blue, linkcolor=blue, pagecolor=blue, citecolor=blue]{hyperref} 

\let\oldcfrac\cfrac
\renewcommand{\cfrac}[2]{\oldcfrac{#1}{#2}\,}

\begin{document}

\title{Derivation of the Kelbg potential/functional}

\author{G. S. Demyanov}
\affiliation{Joint Institute for High Temperatures, Izhorskaya 13 Building 2, Moscow 125412, Russia}
\affiliation{Moscow Institute of Physics and Technology, Institutskiy Pereulok 9, Dolgoprudny, Moscow Region, 141700, Russia}
\author{P. R. Levashov}
\affiliation{Joint Institute for High Temperatures, Izhorskaya 13 Building 2, Moscow 125412, Russia}
\affiliation{Moscow Institute of Physics and Technology, Institutskiy Pereulok 9, Dolgoprudny, Moscow Region, 141700, Russia}

\date{\today}

\begin{abstract}
The density matrix for a system of particles interacting via the Coulomb potential is obtained in the high--temperature limit following almost entirely the original work by Kelbg. For this purpose the Bl\"och equation is solved in the first order of perturbation theory.
We tried to explain all the transformations in the derivation in order to simplify the understanding of this non-trivial theory. The solution of Kelbg is widely used in path integral simulations of Coulomb systems.
\end{abstract}

\maketitle

\section{Introduction}
In 1963 \cite{Kelbg:Annalen:1963} Kelbg calculated the density matrix of a Coulomb system in the first order of perturbation theory. The Kelbg's solution originated a function that resembled some potential and was finite at small distances. This function is often called a ``Kelbg potential'', or ``Kelbg pseudopotential''; the last term is more adequate as the function depends on both distance and temperature. Nevertheless, the Kelbg's solution should be interpreted as some expression for the two--particle density matrix of a Coulomb system at high temperatures. In the case of an arbitrary interparticle potential it is reasonable to call the Kelbg's solution as the Kelbg functional. 

The Kelbg pseudopotential is often used in path integral Monte Carlo (PIMC) calculations \cite{Filinov:PRE:2004, Filinov:IOP:2001, Fraser:PRB:1996, DORNHEIM:PhysRep:2018}. However, the original paper \cite{Kelbg:Annalen:1963} contains only a very brief derivation. Thus, the Kelbg's reasoning is rather difficult to understand. In this report, we formulate a detailed derivation of the Kelbg pseudopotential in the diagonal and non-diagonal cases. We do not produce any new results: our aim is to simplify the understanding of the Kelbg's fundamental work. We follow almost entirely the work \cite{Kelbg:Annalen:1963}. In this paper, we use Mathematica \cite{Mathematica} to calculate integrals.

\section{Primary notations}
We consider $N$ particles interacting through a pair potential $u_{ij}(\textbf{r}_i-\textbf{r}_j)$. The Hamiltonian $\hat{H}$ of the system is:
\begin{equation}
\hat{H} = \hat{K} + \hat{V}, \quad \hat{K} = \sum_{i = 1}^N \cfrac{\hat{\mathbf{p}}_i^2}{2m}, \quad \mathbf{\hat p}_i = -i\hbar \nabla_i,
\end{equation}
\begin{equation}
\label{eq:potentialVop}
\hat{V} = \cfrac{1}{2}\sum_{i=1}^{N}\sum_{\substack{j = 1\\ j\neq i}}^{N}u_{ij}(\textbf{r}_i-\textbf{r}_j),
\end{equation}
where $\hat{\textbf{p}}_i$ is the momentum operator of an $i$-th particle, $m$ is the particle mass, $\textbf{r}_i$ is the coordinate variable of an $i$-th particle. Let us also introduce a variable for the set of all the coordinates:
\begin{equation}
(\textbf{r}_1, \textbf{r}_2, \dots, \textbf{r}_N) \equiv \textbf{R}.
\end{equation}
The time--independent Schr\"{o}dinger equation produces an eigenfunction $\Psi_i(\textbf{R})$ with a corresponding energy value $E_i$:
\begin{equation}
\hat{H} \Psi_i(\textbf{R}) = E_i \Psi_i(\textbf{R}),
\end{equation}
\begin{equation}
\Psi_i(\textbf{R}) \equiv \Psi_i(\textbf{r}_1, \textbf{r}_2, \dots, \textbf{r}_N) = \langle \textbf{R} |\Psi_i\rangle.
\end{equation}
Here, $i$ enumerates the states of the $N$-particle system. We assume that the eigenfunctions $\Psi_i(\textbf{R})$ are orthonormal and form a complete system.

Let us contact the system  with a thermostat with a temperature $T$. We define the density matrix or density operator as follows:
\begin{equation}
\label{eq:densityMatrDef}
\hat{\rho}(\beta) = \exp(-\beta \hat{H}) = \exp(-\beta \hat{K}-\beta \hat{V}),
\end{equation}
where $\beta = 1/(k_B T)$ and $k_B$ is the Boltzmann constant. Note, that we use a non-normalized density matrix. Thus, the partition function $Q(\beta)$ is:
\begin{equation}
Q(\beta) = \mathrm{Sp}\, \hat{\rho}(\beta).
\label{eq:partfunc}
\end{equation}

In the coordinate representation the density matrix $\rho(\textbf{R}, {\textbf{R}'}; \beta)$ has the form:
\begin{equation}
\label{eq:matrixCoordRepres}
\rho(\textbf{R}, {\textbf{R}'}; \beta) = \langle \textbf{R} |\hat{\rho}(\beta)| {\textbf{R}'}\rangle = \sum_ie^{-\beta E_i}\Psi^*_i(\textbf{R})\Psi_i({\textbf{R}'}).
\end{equation}
Here, ${\textbf{R}'}$ denotes the set of primed coordinates ${\textbf{R}'} = (\textbf{r}'_1, \textbf{r}'_2, \dots, \textbf{r}'_N)$ and $\Psi^*_i(\textbf{R})$ is the complex conjugate of $\Psi_i(\textbf{R})$. In Eq. \eqref{eq:matrixCoordRepres} the summation is performed over all states without symmetrization or antisymmetrization. Eq.~\eqref{eq:partfunc} turns into the following form in the coordinate representation:
\begin{equation}
Q(\beta) = \int d\textbf{R}\rho(\textbf{R}, \textbf{R}; \beta), \quad d\textbf{R} \equiv d\textbf{r}_1d\textbf{r}_2\dots d\textbf{r}_N.
\end{equation}

The density matrix satisfies the Bl\"{o}ch equation:
\begin{equation}
\cfrac{d  \hat{\rho}(\beta)}{d  \beta} = - \hat{H}\hat{\rho}(\beta).
\end{equation}

In the further derivation we will often use the ``completeness relation'':
\begin{equation}
\label{eq:coordIdentity}
\hat{1} = \int d\textbf{R} |\textbf{R}\rangle\langle \textbf{R}|.
\end{equation}

Two conjugate variables $\textbf{G}\equiv (\textbf{g}_1, \textbf{g}_2, \dots, \textbf{g}_N)$ and $\textbf{R} \equiv (\textbf{r}_1, \textbf{r}_2, \dots, \textbf{r}_N)$ satisfy the following identities:
\begin{equation}
\langle \textbf{R} | \textbf{G} \rangle \equiv  \langle \textbf{r}_1, \textbf{r}_2, \dots, \textbf{r}_N| \textbf{g}_1, \textbf{g}_2, \dots, \textbf{g}_N \rangle = \cfrac{1}{(2\pi\hbar)^{3N/2}}\, e^{\tfrac{i}{\hbar} \textbf{R} \cdot \textbf{G} }.
\end{equation}
\begin{equation}
\textbf{R}\cdot \textbf{G} \equiv \textbf{r}_1 \textbf{g}_1 + \textbf{r}_2 \textbf{g}_2+\dots+\textbf{r}_N \textbf{g}_N.
\end{equation}
An arbitrary variable $\mathbf{Q} = (\mathbf{q}_1,\ldots, \mathbf{q}_N)$ will be used below, where $\mathbf{q}_i$ is the coordinate of an $i$-th particle. 

The action of the momentum operator $\hat{\textbf{p}}_i$ on $| \textbf{G} \rangle$ defines the momentum variable $\textbf{g}_i$:
\begin{equation}
\hat{\textbf{p}}_i| \textbf{G} \rangle = \textbf{g}_i| \textbf{G} \rangle.
\end{equation}
Another ``completeness relation'' for the conjugate variable $\textbf{G}$ will be used:
\begin{equation}
\label{eq:conjcompl}
\hat{1} = \int\cfrac{d\textbf{G}}{(2\pi\hbar)^{3N/2}} \, |\textbf{G}\rangle\langle\textbf{G} |.
\end{equation}

\section{Derivation of the Kelbg functional}
Our goal is to separate the kinetic and potential energy in Eq. \eqref{eq:densityMatrDef}. 
Kelbg \cite{Kelbg:Annalen:1963} introduces a correction function  (``Korrekturfunktion''), which we write as an operator $\hat G(\beta)$:
\begin{equation}
\label{eq:Gdef}
\exp(-\beta(\hat{K} + \hat{V})) = \exp(-\beta\hat{V})\exp(-\beta\hat{K}) \hat{G}(\beta).
\end{equation}
The Baker–Campbell–Hausdorff (BCH) relation gives the exact formula for $\hat{G}(\beta)$:
\begin{equation}
\label{eq:BCH}
\hat{G}(\beta) = \hat{1} \exp\left(-\cfrac{\beta^2}{2}\, [\hat{V}, \hat{K}]\right)
\exp\left(\cfrac{\beta^3}{12}\, ([\hat{V},[\hat{V}, \hat{K}]] + [\hat{K},[\hat{K},\hat{V}]])\right)\dots,
\end{equation}
where $[\hat{V}, \hat{K}]$ is a commutator of $\hat{V}$ and $\hat{K}$. Our goal is to consider $\hat{G}(\beta)$ in the first order of $\hat{V}$. We cannot do it directly from Eq. \eqref{eq:BCH}: the first-order contributions are included not only in the first exponent ($\beta^2$), but also, for example, in the second one ($\beta^3$).

Therefore, we (according to Kelbg) differentiate Eq. \eqref{eq:Gdef} by $\beta$:
\begin{equation}
\label{eq:Gdiffbeta}
-(\hat{K} + \hat{V})e^{-\beta(\hat{K} + \hat{V})} = -\hat{V}e^{-\beta\hat{V}}e^{-\beta\hat{K}}\hat{G}(\beta)-e^{-\beta\hat{V}}\hat{K}e^{-\beta\hat{K}}\hat{G}(\beta)
+
e^{-\beta\hat{V}}e^{-\beta\hat{K}}\cfrac{d  \hat{G}(\beta) }{d  \beta}.
\end{equation}
Then we rewrite the left side of Eq. \eqref{eq:Gdiffbeta} using \eqref{eq:Gdef}:
\begin{multline}
\label{eq:dghf}
-(\hat{K} + \hat{V})e^{-\beta(\hat{K} + \hat{V})} = -(\hat{K} + \hat{V})\exp(-\beta\hat{V})\exp(-\beta\hat{K}) \hat{G}(\beta) \\=
-\hat{K}\exp(-\beta\hat{V})\exp(-\beta\hat{K}) \hat{G}(\beta)
-\hat{V}\exp(-\beta\hat{V})\exp(-\beta\hat{K}) \hat{G}(\beta).
\end{multline}
After substituting Eq. \eqref{eq:dghf} in Eq. \eqref{eq:Gdiffbeta} and eliminating the same terms, we get:
\begin{equation}
\label{eq:gfdgeg}
e^{-\beta\hat{V}}e^{-\beta\hat{K}}\cfrac{d  \hat{G}(\beta) }{d  \beta}=
e^{-\beta\hat{V}}\hat{K}e^{-\beta\hat{K}}\hat{G}(\beta)
-
\hat{K}e^{-\beta\hat{V}}e^{-\beta\hat{K}} \hat{G}(\beta) = e^{-\beta\hat{V}} \left(\hat{K} - e^{\beta\hat{V}}\hat{K}e^{-\beta\hat{V}}\right)e^{-\beta\hat{K}}\hat{G}(\beta).
\end{equation}
Multiplying Eq. \eqref{eq:gfdgeg} from the left by 
$e^{\beta\hat{K}}e^{\beta\hat{V}}$, we obtain \cite[Eq. (11)]{Kelbg:Annalen:1963}:
\begin{equation}
\label{eq:11Eq}
\cfrac{d  \hat{G}(\beta) }{d  \beta}=e^{\beta\hat{K}} \left(\hat{K} - e^{\beta\hat{V}}\hat{K}e^{-\beta\hat{V}}\right)e^{-\beta\hat{K}}\hat{G}(\beta).
\end{equation}

Now we can use the series expansion:
\begin{equation}
e^{\beta \hat{V}} = \hat{1} + \beta \hat{V} + \cfrac{\beta^2}{2} \hat{V}^2 + \dots, \quad e^{-\beta \hat{V}} = \hat{1} - \beta \hat{V} + \cfrac{\beta^2}{2} \hat{V}^2 + \dots 
\end{equation}
So the second term in brackets in Eq. \eqref{eq:11Eq} becomes:
\begin{equation}
\label{eq:series123}
e^{\beta\hat{V}}\hat{K}e^{-\beta\hat{V}} = \hat{K} +\beta[\hat{V},\hat{K}]+\cfrac{\beta^2}{2}[\hat{V},[\hat{V},\hat{K}]]+\cfrac{\beta^3}{6}[\hat{V},[\hat{V},[\hat{V},\hat{K}]]] + \ldots
\end{equation}
Substituting Eq. \eqref{eq:series123} in Eq. \eqref{eq:11Eq}, we obtain a series with nested commutators \cite[Eq. (12)]{Kelbg:Annalen:1963}:
\begin{equation}
\label{eq:linearGdiff}
\cfrac{d  \hat{G}(\beta) }{d  \beta}=- e^{\beta\hat{K}}\left\{\beta[\hat{V},\hat{K}]+\cfrac{\beta^2}{2}[\hat{V},[\hat{V},\hat{K}]]+\cfrac{\beta^3}{6}[\hat{V},[\hat{V},[\hat{V},\hat{K}]]] + \ldots\right\}e^{-\beta\hat{K}}\hat{G}(\beta).
\end{equation}
Kelbg states, that ``the series breaks off with the term $\beta^2$ because of the structure of the kinetic energy operator''. He gives no proof of the series truncation. 
Nevertheless, this statement is true since the kinetic energy operator contains only second order derivatives. Thus, the commutator $[\hat{V},[\hat{V},[\hat{V},\hat{K}]]] = 0$ (see App. \ref{sec:app1}) that leads to zeroing of higher $\beta$-order terms and \cite[Eq. (12)]{Kelbg:Annalen:1963} is correct and \emph{exact}. 
The term of order $\beta^2$ in Eq. \eqref{eq:linearGdiff} is not used further anyway.

Next, we use the following transformation:
\begin{equation}
\label{eq:poysn1}
e^{\beta\hat{K}}[\hat{K}, \hat{V}]e^{-\beta\hat{K}}
=
\left(
\hat{K}e^{\beta\hat{K}}\hat{V}e^{-\beta\hat{K}}
-
e^{\beta\hat{K}}\hat{V}\hat{K}e^{-\beta\hat{K}}
\right)
=
 \cfrac{d }{d  \beta}\left(e^{\beta\hat{K}}\hat{V}e^{-\beta\hat{K}}\right)
\end{equation}
to rewrite Eq. \eqref{eq:linearGdiff} in the following form (in the first order of $\hat{V}$):
\begin{equation}
\label{eq:GdiffLinear}
\cfrac{d  \hat{G}(\beta) }{d  \beta}=-\beta e^{\beta\hat{K}}[\hat{V},\hat{K}]e^{-\beta\hat{K}}\hat{G}(\beta)
=
\beta \cfrac{d }{d  \beta}\left(e^{\beta\hat{K}}\hat{V}e^{-\beta\hat{K}}\right)\hat{G}(\beta).
\end{equation}
From the definition \eqref{eq:Gdef}:
\begin{equation}
\hat{G}(0) = \hat{1}.
\end{equation}
Providing the formal integration of Eq.  \eqref{eq:GdiffLinear} over $\beta$, we obtain \cite[Eq. (13)]{Kelbg:Annalen:1963}:
\begin{multline}
\hat{G}(\beta) = \hat{1} + \int\limits_0^{\beta}\beta_1 \cfrac{d }{d  \beta_1}\left(e^{\beta_1\hat{K}}\hat{V}e^{-\beta_1\hat{K}}\right) \hat{G}(\beta)d\beta_1 = \hat{1} + \int\limits_0^{\beta}\beta_1 \cfrac{d }{d  \beta_1}\left(e^{\beta_1\hat{K}}\hat{V}e^{-\beta_1\hat{K}}\right) \hat{1}d\beta_1 
\\+ 
\int\limits_0^{\beta}\beta_1 \cfrac{d }{d  \beta_1}\left(e^{\beta_1\hat{K}}\hat{V}e^{-\beta_1\hat{K}}\right) \int\limits_0^{\beta_1}\beta'_1 \cfrac{d }{d  \beta'_1}\left(e^{\beta'_1\hat{K}}\hat{V}e^{-\beta'_1\hat{K}}\right) \hat{1}d\beta'_1d\beta_1 + $\ldots$.
\end{multline}
Since we are looking for $\hat{G}(\beta)$ in the first order of perturbation theory, we omit all the terms except for the first two:
\begin{equation}
\label{eq:Gdecision}
\hat{G}(\beta)  = \hat{1} + \int\limits_0^{\beta}\beta_1 \cfrac{d }{d  \beta_1}\left(e^{\beta_1\hat{K}}\hat{V}e^{-\beta_1\hat{K}}\right)d\beta_1.
\end{equation}
Substituting Eq. \eqref{eq:Gdecision} into Eq. \eqref{eq:Gdef}, we get the density matrix:
\begin{equation}
\label{eq:matriPlot1}
\hat{\rho}(\beta) = e^{-\beta\hat{V}}e^{-\beta\hat{K}} + e^{-\beta\hat{V}}e^{-\beta\hat{K}}\int\limits_0^{\beta}\beta_1 \cfrac{d }{d  \beta_1}\left(e^{\beta_1\hat{K}}\hat{V}e^{-\beta_1\hat{K}}\right) d\beta_1.
\end{equation}
This is the general equation in \cite{Kelbg:Annalen:1963}. In the rest of the article we will transform it to different forms.

\subsection{Transformation of density operator Eq. \eqref{eq:matriPlot1}}
First, we transform the integral term in Eq. \eqref{eq:matriPlot1} by differentiating it over $\beta_1$ and multiplying the integrand by $e^{-\beta\hat{K}}$:
\begin{multline}
e^{-\beta\hat{V}}e^{-\beta\hat{K}}\int\limits_0^{\beta}\beta_1 \cfrac{d }{d  \beta_1}\left(e^{\beta_1\hat{K}}\hat{V}e^{-\beta_1\hat{K}}\right) d\beta_1=
e^{-\beta\hat{V}}e^{-\beta\hat{K}}\int\limits_0^{\beta}\beta_1 e^{\beta_1\hat{K}}\left(
\hat{K}\hat{V} - \hat{V}\hat{K}
\right)e^{-\beta_1\hat{K}}\  \hat{1} \ d\beta_1
\\=
e^{-\beta\hat{V}}
\int\limits_0^{\beta}\beta_1 e^{(\beta_1-\beta)\hat{K}}\left(
\hat{K}\hat{V} - \hat{V}\hat{K}
\right)e^{-(\beta_1-\beta)\hat{K}} e^{-\beta\hat{K}}d\beta_1.
\end{multline}
We also used here, that $\hat{1} = e^{\beta\hat{K}}e^{-\beta\hat{K}}$. Now we will again transform this equation to the form of a derivative:
\begin{equation}
\int\limits_0^{\beta}\beta_1 e^{(\beta_1-\beta)\hat{K}}\left(
\hat{K}\hat{V} - \hat{V}\hat{K}
\right)e^{-(\beta_1-\beta)\hat{K}} e^{-\beta\hat{K}}d\beta_1
=
\int\limits_0^{\beta}\beta_1 \cfrac{d }{d  \beta_1}\left(e^{(\beta_1-\beta)\hat{K}}\hat{V}e^{-(\beta_1-\beta)\hat{K}}\right) e^{-\beta\hat{K}}d\beta_1.
\end{equation}
Finally, Eq.~\eqref{eq:matriPlot1} is transformed into \cite[Eq. (14)]{Kelbg:Annalen:1963}:
\begin{equation}
\label{eq:eq14}
\hat{\rho}(\beta) = e^{-\beta\hat{V}}e^{-\beta\hat{K}} + e^{-\beta\hat{V}}\int\limits_0^{\beta}\beta_1 \cfrac{d }{d  \beta_1}\left(e^{(\beta_1-\beta)\hat{K}}\hat{V}e^{-(\beta_1-\beta)\hat{K}}\right) e^{-\beta\hat{K}}d\beta_1.
\end{equation}
Note that the factor $e^{\beta\hat K}$ can be introduced under the derivative, since it does not depend on $\beta_1$:
\begin{equation}
\cfrac{d }{d  \beta_1}\left(e^{(\beta_1-\beta)\hat{K}}\hat{V}e^{-(\beta_1-\beta)\hat{K}}\right) e^{-\beta\hat{K}}
=
\cfrac{d }{d  \beta_1}\left(e^{(\beta_1-\beta)\hat{K}}\hat{V}e^{-(\beta_1-\beta)\hat{K}}e^{-\beta\hat{K}}\right).
\end{equation}

Following Kelbg, we rewrite the inter-particle potential through the parameters $e_i$ and $D$:
\begin{equation}
u_{ij}(\textbf{r}_i-\textbf{r}_j) = \cfrac{e_ie_j}{D}\,u(\textbf{r}_i-\textbf{r}_j).
\end{equation}
We suppose that $e_i$ has the meaning of a charge and $D$ of a length. We decompose $u(\textbf{r}_i-\textbf{r}_j)$ into a Fourier integral:
\begin{equation}
u(\textbf{r}_i-\textbf{r}_j) = \cfrac{1}{(2\pi)^3}\int d\textbf{t} v(\textbf{t})e^{i\textbf{t}(\textbf{r}_i-\textbf{r}_j)}.
\end{equation}
Here, $v(\textbf{t})$ is a Fourier component of the potential $u(\textbf{r}_i-\textbf{r}_j)$. Then the full potential energy has the following form:
\begin{equation}
\hat{V} = \cfrac{1}{16\pi^3D}\sum_{i=1}^{N}\sum_{\substack{j = 1\\ j\neq i}}^{N}e_ie_j\int d\textbf{t} v(\textbf{t})e^{i\textbf{t}(\textbf{r}_i-\textbf{r}_j)}.
\end{equation}
Thus Eq. \eqref{eq:eq14} becomes:
\begin{equation}
\label{eq:urrhoint}
\hat{\rho}(\beta) = e^{-\beta\hat{V}}e^{-\beta\hat{K}} + 
\cfrac{1}{16\pi^3D}\sum_{i=1}^{N}\sum_{\substack{j = 1\\ j\neq i}}^{N}e_ie_j\int d\textbf{t} v(\textbf{t})
e^{-\beta\hat{V}}\int\limits_0^{\beta}\beta_1 \cfrac{d }{d  \beta_1}
\left(e^{(\beta_1-\beta)\hat{K}}
e^{i\textbf{t}(\textbf{r}_i-\textbf{r}_j)}
e^{-(\beta_1-\beta)\hat{K}}\right) e^{-\beta\hat{K}}d\beta_1.
\end{equation}

We introduce the following notation \cite[Eq. (17)]{Kelbg:Annalen:1963}:
\begin{equation}
\label{eq:quest1F}
\hat{F}_{ij} = e^{\beta'\hat{K}}
e^{i\textbf{t}(\textbf{r}_i-\textbf{r}_j)}
e^{-\beta'\hat{K}}, \quad \beta' = \beta_1-\beta.
\end{equation}
$\hat F_{ij}$ corresponds to the expression under the derivative in Eq. \eqref{eq:urrhoint}. We are going to calculate the action of $e^{\beta'\hat{K}}$  on $e^{i\textbf{t}(\textbf{r}_i-\textbf{r}_j)}$. If $f(x)$ is some function, then:
\begin{equation}
\label{eq:funcActionDef}
f(\hat{H})\Psi_i(\textbf{R}) = f(E_i)\Psi_i(\textbf{R}).
\end{equation}
For the case of $\hat{K}$, we consider the action of $\hat{K}$ on $e^{i\textbf{t}(\textbf{r}_i-\textbf{r}_j)}$:
\begin{equation}
\label{eq:sdhjfdsh}
\hat{K}e^{i\textbf{t}(\textbf{r}_i-\textbf{r}_j)} = \cfrac{1}{2m} \sum_{n = 1}^N \hat{\textbf{p}}^2_n e^{i\textbf{t}(\textbf{r}_i-\textbf{r}_j)} = 
\cfrac{1}{2m} (\hat{\textbf{p}}^2_i+\hat{\textbf{p}}^2_j) e^{i\textbf{t}(\textbf{r}_i-\textbf{r}_j)} +\cfrac{e^{i\textbf{t}(\textbf{r}_i-\textbf{r}_j)}}{2m} \sum_{\substack{n = 1\\ n\neq i,j}}^N \hat{\textbf{p}}^2_n
\end{equation}
to calculate the action of $e^{\beta'\hat{K}}$  on $e^{i\textbf{t}(\textbf{r}_i-\textbf{r}_j)}$.
Let us consider the actions of individual contributions in Eq. \eqref{eq:sdhjfdsh} on the wave-function $\Psi(\textbf{R})$:
\begin{equation}
\hat{\textbf{p}}^2_i e^{i\textbf{t}(\textbf{r}_i-\textbf{r}_j)} \Psi(\textbf{R}) = 
-\hbar^2 \nabla_i^2e^{i\textbf{t}(\textbf{r}_i-\textbf{r}_j)} \Psi(\textbf{R}).
\end{equation}
The actions of $\nabla_i^2$ and $\nabla_j^2$ produces three terms:
\begin{equation}
\label{eq:nablaSqContri}
\nabla_i^2e^{i\textbf{t}(\textbf{r}_i-\textbf{r}_j)} \Psi(\textbf{R})
=
-t^2 e^{i\textbf{t}(\textbf{r}_i-\textbf{r}_j)} \Psi(\textbf{R}) +  2 i e^{i\textbf{t}(\textbf{r}_i-\textbf{r}_j)} \textbf{t}  \nabla_i \Psi(\textbf{R}) + 
e^{i\textbf{t}(\textbf{r}_i-\textbf{r}_j)} \nabla_i^2  \Psi(\textbf{R}),
\end{equation}
\begin{equation}
\label{eq:nablaSqContrj}
\nabla_j^2e^{i\textbf{t}(\textbf{r}_i-\textbf{r}_j)} \Psi(\textbf{R})
=
-t^2 e^{i\textbf{t}(\textbf{r}_i-\textbf{r}_j)} \Psi(\textbf{R}) -  
2 i e^{i\textbf{t}(\textbf{r}_i-\textbf{r}_j)} \textbf{t}  \nabla_j \Psi(\textbf{R}) + 
e^{i\textbf{t}(\textbf{r}_i-\textbf{r}_j)} \nabla_j^2  \Psi(\textbf{R}).
\end{equation}
Summing it all up, we obtain:
\begin{equation}
\hat{K}e^{i\textbf{t}(\textbf{r}_i-\textbf{r}_j)} \Psi(\textbf{R}) =
\left(\cfrac{\hbar^2t^2}{m}  e^{i\textbf{t}(\textbf{r}_i-\textbf{r}_j)} +  e^{i\textbf{t}(\textbf{r}_i-\textbf{r}_j)} \cfrac{\hbar}{m}  \textbf{t} (\hat{\textbf{p}}_i - \hat{\textbf{p}}_j)\right)\Psi(\textbf{R})
+
e^{i\textbf{t}(\textbf{r}_i-\textbf{r}_j)} \cfrac{1}{2m}  \sum_{n = 1}^N \hat{\textbf{p}}^2_n\Psi(\textbf{R}),
\end{equation}
and in the operator notation:
\begin{equation}
\hat{K}e^{i\textbf{t}(\textbf{r}_i-\textbf{r}_j)} =
e^{i\textbf{t}(\textbf{r}_i-\textbf{r}_j)}\left(\cfrac{\hbar^2t^2}{m}   +  \cfrac{\hbar}{m}  \textbf{t} (\hat{\textbf{p}}_i - \hat{\textbf{p}}_j)
+
\hat{K}\right).
\end{equation}
Now we can compute the action of $e^{\beta'\hat{K}}$ on $e^{i\textbf{t}(\textbf{r}_i-\textbf{r}_j)}$ similarly to Eq. \eqref{eq:funcActionDef}:
\begin{equation}
e^{\beta'\hat{K}}e^{i\textbf{t}(\textbf{r}_i-\textbf{r}_j)} =
e^{i\textbf{t}(\textbf{r}_i-\textbf{r}_j)}
 e^{\beta' \tfrac{\hbar^2t^2}{m} + \beta'\tfrac{\hbar}{m}  \textbf{t} (\hat{\textbf{p}}_i - \hat{\textbf{p}}_j)} e^{\beta'\hat{K}}.
\end{equation}
Thus, we obtain \cite[Eq. (18)]{Kelbg:Annalen:1963}:
\begin{equation}
\label{Fijderiv}
\hat{F}_{ij} = e^{\beta'\hat{K}}
e^{i\textbf{t}(\textbf{r}_i-\textbf{r}_j)}
e^{-\beta'\hat{K}} 
= e^{i\textbf{t}(\textbf{r}_i-\textbf{r}_j)}
e^{\beta' \tfrac{\hbar^2t^2}{m} + \beta'\tfrac{\hbar}{m}  \textbf{t} (\hat{\textbf{p}}_i - \hat{\textbf{p}}_j)} e^{\beta'\hat{K}}e^{-\beta'\hat{K}} 
=
e^{i\textbf{t}(\textbf{r}_i-\textbf{r}_j)}e^{\beta' \tfrac{\hbar^2t^2}{m} + \beta'\tfrac{\hbar}{m}  \textbf{t} (\hat{\textbf{p}}_i - \hat{\textbf{p}}_j)}.
\end{equation}

Now we substitute Eq. \eqref{Fijderiv} in Eq. \eqref{eq:urrhoint}:
\begin{multline}
\label{eq:ghewjd}
\hat{\rho}(\beta) = e^{-\beta\hat{V}}e^{-\beta\hat{K}} \\ 
{} + \cfrac{1}{16\pi^3D}\sum_{i=1}^{N}\sum_{\substack{j = 1\\ j\neq i}}^{N}e_ie_j 
\int d\textbf{t} v(\textbf{t}) e^{i\textbf{t}(\textbf{r}_i-\textbf{r}_j)}
e^{-\beta\hat{V}}\int\limits_0^{\beta}\beta_1 \cfrac{d }{d  \beta_1}
\left(
e^{\tfrac{\hbar (\beta_1-\beta)}{m}\,  \textbf{t} (\hat{\textbf{p}}_i - \hat{\textbf{p}}_j) +  \tfrac{\hbar^2(\beta_1-\beta)}{m}\,t^2}
\right) 
e^{-\beta\hat{K}}d\beta_1
\end{multline}
to obtain \cite[Eq. (19)]{Kelbg:Annalen:1963}. 

\subsection{Coordinate representation of density matrix}

Now we are going to calculate the density matrix \eqref{eq:ghewjd} in the coordinate representation, that is  $\langle \textbf{R} |\hat{\rho}(\beta)| {\textbf{R}'}\rangle$.
To do this, we should calculate the following matrix element: 
\begin{equation}
\label{eq:ujsdbvuhi}
\langle \textbf{R} |e^{-\beta\hat{V}}e^{\tfrac{\hbar (\beta_1-\beta)}{m}\,  \textbf{t} (\hat{\textbf{p}}_i - \hat{\textbf{p}}_j)}e^{-\beta\hat{K}}| {\textbf{R}'}\rangle.
\end{equation}
First we will insert a coordinate variable $\textbf{Q} \equiv (\textbf{q}_1, \textbf{q}_2, \dots, \textbf{q}_N)$, using Eq. \eqref{eq:coordIdentity}:
\begin{equation}
\langle \textbf{R} |e^{-\beta\hat{V}}
\hat{1}
e^{\tfrac{\hbar (\beta_1-\beta)}{m}\,  \textbf{t} (\hat{\textbf{p}}_i - \hat{\textbf{p}}_j)}e^{-\beta\hat{K}}| {\textbf{R}'}\rangle
=
\int d\textbf{Q}
\langle \textbf{R} |e^{-\beta\hat{V}}|\textbf{Q}\rangle
\langle \textbf{Q}|e^{\tfrac{\hbar (\beta_1-\beta)}{m}\,  \textbf{t} (\hat{\textbf{p}}_i - \hat{\textbf{p}}_j)}e^{-\beta\hat{K}}| {\textbf{R}'}\rangle.
\end{equation}
To calculate $\langle \textbf{R} |e^{-\beta\hat{V}}|\textbf{Q}\rangle$ we note that:
\begin{equation}
\langle\textbf{R} |\hat{V}| \textbf{Q}\rangle
=
U(\textbf{R})\delta(\textbf{R}-\textbf{Q}),
\end{equation}
where $U(\textbf{R})$ is a potential energy (function, not an operator).  Thus:
\begin{equation}
\label{eq:pot_matrx}
\langle \textbf{R} |e^{-\beta\hat{V}}|\textbf{Q}\rangle = e^{-\beta U(\textbf{R})} \delta(\textbf{R}-\textbf{Q}).
\end{equation}
So Eq. \eqref{eq:ujsdbvuhi} transforms into:
\begin{equation}
\langle \textbf{R} |e^{-\beta\hat{V}}e^{\tfrac{\hbar (\beta_1-\beta)}{m}\,  \textbf{t} (\hat{\textbf{p}}_i - \hat{\textbf{p}}_j)}e^{-\beta\hat{K}}| {\textbf{R}'}\rangle
=
e^{-\beta U(\textbf{R}) }
\langle \textbf{R} |e^{\tfrac{\hbar (\beta_1-\beta)}{m}\,  \textbf{t} (\hat{\textbf{p}}_i - \hat{\textbf{p}}_j)}e^{-\beta\hat{K}}| {\textbf{R}'}\rangle.
\end{equation}

Now we are going to calculate $\langle \textbf{R} |e^{\tfrac{\hbar (\beta_1-\beta)}{m}\,  \textbf{t} (\hat{\textbf{p}}_i - \hat{\textbf{p}}_j)}e^{-\beta\hat{K}}| {\textbf{R}'}\rangle$. For this purpose we insert one more coordinate variable, using Eq. \eqref{eq:coordIdentity}:
\begin{equation}
\label{eq:ddfgrgf}
\langle \textbf{R} |e^{\tfrac{\hbar (\beta_1-\beta)}{m}\,  \textbf{t} (\hat{\textbf{p}}_i - \hat{\textbf{p}}_j)}
\hat{1}
e^{-\beta\hat{K}}| {\textbf{R}'}\rangle
=
\int d\textbf{Q}\langle \textbf{R} |e^{\tfrac{\hbar (\beta_1-\beta)}{m}\,  \textbf{t} (\hat{\textbf{p}}_i - \hat{\textbf{p}}_j)}| {\textbf{Q}}\rangle 
\langle \textbf{Q} |e^{-\beta\hat{K}}| {\textbf{R}'}\rangle.
\end{equation}
Consider the first matrix element $\langle \textbf{R} |e^{\tfrac{\hbar (\beta_1-\beta)}{m}\,  \textbf{t} (\hat{\textbf{p}}_i - \hat{\textbf{p}}_j)}| {\textbf{Q}}\rangle$ in Eq. \eqref{eq:ddfgrgf}. Let us insert a \emph{conjugate} variable $\textbf{G}$, using Eq. \eqref{eq:conjcompl}:
\begin{equation}
\langle \textbf{R} |e^{\tfrac{\hbar (\beta_1-\beta)}{m}\,  \textbf{t} (\hat{\textbf{p}}_i - \hat{\textbf{p}}_j)}| {\textbf{Q}}\rangle 
=
\langle \textbf{R} |e^{\tfrac{\hbar \beta'}{m}\,  \textbf{t} \hat{\textbf{p}}_i} 
\hat{1}
e^{-\tfrac{\hbar \beta'}{m}\,  \textbf{t} \hat{\textbf{p}}_j}| {\textbf{Q}}\rangle 
=
\int \cfrac{d\textbf{G}}{(2\pi\hbar)^{3N/2}} \,
\langle \textbf{R} |e^{\tfrac{\hbar \beta'}{m}\,  \textbf{t} \hat{\textbf{p}}_i} |\textbf{G}\rangle
\langle\textbf{G} |e^{-\tfrac{\hbar \beta'}{m}\,  \textbf{t} \hat{\textbf{p}}_j}| {\textbf{Q}}\rangle. 
\end{equation}
The momentum operator is Hermitian; it acts on the ket-vector in the first case and on the bra-vector in the second:
\begin{equation}
\langle \textbf{R} |e^{\tfrac{\hbar \beta'}{m}\,  \textbf{t} \hat{\textbf{p}}_i} |\textbf{G}\rangle = e^{\tfrac{\hbar \beta'}{m}\,  \textbf{t} \textbf{g}_i}\langle \textbf{R} |\textbf{G}\rangle,
\end{equation}
\begin{equation}
\langle\textbf{G} |e^{-\tfrac{\hbar \beta'}{m}\,  \textbf{t} \hat{\textbf{p}}_j}| {\textbf{Q}}\rangle = e^{-\tfrac{\hbar \beta'}{m}\,  \textbf{t} \textbf{g}_j}\langle\textbf{G} | {\textbf{Q}}\rangle.
\end{equation}
Finally, we get:
\begin{multline}
\label{eq:sdfsdf}
\langle \textbf{R} |e^{\tfrac{\hbar (\beta_1-\beta)}{m}\,  \textbf{t} (\hat{\textbf{p}}_i - \hat{\textbf{p}}_j)}| {\textbf{Q}}\rangle 
=
\int \cfrac{d\textbf{G}}{(2\pi\hbar)^{3N/2}} \, e^{\tfrac{\hbar \beta'}{m}\,  \textbf{t} (\textbf{g}_i-\textbf{g}_j)}\langle \textbf{R} |\textbf{G}\rangle\langle\textbf{G} | {\textbf{Q}}\rangle \\
{} =
\int \cfrac{d\textbf{G}}{(2\pi\hbar)^{3N}} \,e^{\tfrac{\hbar \beta'}{m}\,  \textbf{t} (\textbf{g}_i-\textbf{g}_j)}
e^{\tfrac{i}{\hbar} \textbf{R} \cdot \textbf{G} }
e^{-\tfrac{i}{\hbar} \textbf{Q}\cdot \textbf{G} }.
\end{multline}
Since
\begin{equation}
e^{\tfrac{i}{\hbar} \textbf{R} \cdot \textbf{G} }
e^{-\tfrac{i}{\hbar} \textbf{Q}\cdot \textbf{G} } = 
e^{\tfrac{i}{\hbar} (\textbf{R} -\textbf{Q})\cdot \textbf{G} }
=
e^{\tfrac{i}{\hbar}
\sum\limits_{m = 1}^N
(\textbf{r}_m -\textbf{q}_m)\cdot \textbf{g}_m
}
=
\prod\limits_{m=1}^{N}
e^{\tfrac{i}{\hbar}
(\textbf{r}_m -\textbf{q}_m)\cdot \textbf{g}_m
}, \quad d\textbf{G} = \prod\limits_{n=1}^{N}d\textbf{g}_n,
\end{equation}
the integral in \eqref{eq:sdfsdf} is easily taken over all variables, except $\textbf{g}_i,\textbf{g}_j$:
\begin{multline}
\label{eq:fdsjksdfhds}
\int \cfrac{d\textbf{G}}{(2\pi\hbar)^{3N}} \,e^{\tfrac{\hbar \beta'}{m}\,  \textbf{t} (\textbf{g}_i-\textbf{g}_j)}
e^{\tfrac{i}{\hbar} \textbf{R} \cdot \textbf{G} }
e^{-\tfrac{i}{\hbar} \textbf{Q}\cdot \textbf{G} }
=
\int \cfrac{\prod\limits_{n=1}^{N}d\textbf{g}_n}{(2\pi\hbar)^{3N}} \,e^{\tfrac{\hbar \beta'}{m}\,  \textbf{t} (\textbf{g}_i-\textbf{g}_j)}
\prod\limits_{m=1}^{N}
e^{\tfrac{i}{\hbar}
(\textbf{r}_m -\textbf{q}_m)\cdot \textbf{g}_m
}
\\ =
\cfrac{1}{(2\pi\hbar)^{3N}} \,
\left(\prod\limits_{\substack{n=1 \\ n\neq i,j}}^{N}\int d\textbf{g}_n e^{\tfrac{i}{\hbar}
(\textbf{r}_n -\textbf{q}_n)\cdot \textbf{g}_n}\right) 
\times 
\left(
\int d\textbf{g}_ie^{\tfrac{i}{\hbar}
(\textbf{r}_i -\textbf{q}_i)\cdot \textbf{g}_i
}
e^{\tfrac{\hbar \beta'}{m}\,  \textbf{t} \textbf{g}_i}
\right)
\times
\left(\int d\textbf{g}_je^{\tfrac{i}{\hbar}
(\textbf{r}_j -\textbf{q}_j)\cdot \textbf{g}_j
}
e^{-\tfrac{\hbar \beta'}{m}\,  \textbf{t} \textbf{g}_j}
\right).
\end{multline}
The first integral is a Dirac $\delta$-function:
\begin{equation}
\int d\textbf{g}_n e^{\tfrac{i}{\hbar}
(\textbf{r}_n -\textbf{q}_n)\cdot \textbf{g}_n} = (2\pi\hbar)^3\delta(\textbf{r}_n -\textbf{q}_n).
\label{eq:rnqn}
\end{equation}
The following two integrals are $\delta$-functions too:
\begin{equation}
\int d\textbf{g}_ie^{\tfrac{i}{\hbar}
(\textbf{r}_i -\textbf{q}_i)\cdot \textbf{g}_i
}
e^{\tfrac{\hbar \beta'}{m}\,  \textbf{t} \textbf{g}_i}
=
\int d\textbf{g}_ie^{\tfrac{i}{\hbar}
(\textbf{r}_i -\textbf{q}_i -i \tfrac{\hbar^2 \beta'}{m}\,  \textbf{t})\cdot \textbf{g}_i}
=
(2\pi\hbar)^3\delta(\textbf{r}_i  -i \tfrac{\hbar^2 \beta'}{m}\,  \textbf{t}-\textbf{q}_i),
\end{equation}
\begin{equation}
\int d\textbf{g}_je^{\tfrac{i}{\hbar}
(\textbf{r}_j -\textbf{q}_j)\cdot \textbf{g}_j
}
e^{-\tfrac{\hbar \beta'}{m}\,  \textbf{t} \textbf{g}_j}
=
(2\pi\hbar)^3\delta(\textbf{r}_j  +i \tfrac{\hbar^2 \beta'}{m}\,  \textbf{t}-\textbf{q}_j).
\label{eq:rjqj}
\end{equation}
Now we substitute \eqref{eq:rnqn}--\eqref{eq:rjqj} in \eqref{eq:fdsjksdfhds} to calculate \eqref{eq:sdfsdf}.
Finally, we have calculated the first matrix element in \eqref{eq:ddfgrgf}:
\begin{equation}
\langle \textbf{R} |e^{\tfrac{\hbar (\beta_1-\beta)}{m}\,  \textbf{t} (\hat{\textbf{p}}_i - \hat{\textbf{p}}_j)}| {\textbf{Q}}\rangle 
=
\left(\prod\limits_{\substack{n=1 \\ n\neq i,j}}\delta(\textbf{r}_n -\textbf{q}_n)\right)
\delta\left((\textbf{r}_i-i \tfrac{\hbar^2 \beta'}{m}\,  \textbf{t})  -\textbf{q}_i\right)
\delta\left((\textbf{r}_j+i \tfrac{\hbar^2 \beta'}{m}\,  \textbf{t})  -\textbf{q}_j\right).
\end{equation}
Substituting it in Eq. \eqref{eq:ddfgrgf} and providing integration over all $\textbf{q}_1\dots \textbf{q}_N$, we obtain:
\begin{multline}
\label{eq:dfhbej}
\langle \textbf{R} |e^{\tfrac{\hbar (\beta_1-\beta)}{m}\,  \textbf{t} (\hat{\textbf{p}}_i - \hat{\textbf{p}}_j)}e^{-\beta\hat{K}}| {\textbf{R}'}\rangle \\ =
\int d\textbf{Q}
\left(\prod\limits_{\substack{n=1 \\ n\neq i,j}}\delta(\textbf{r}_n -\textbf{q}_n)\right)
\delta\left((\textbf{r}_i-i \tfrac{\hbar^2 \beta'}{m}\,  \textbf{t})  -\textbf{q}_i\right)
\delta\left((\textbf{r}_j+i \tfrac{\hbar^2 \beta'}{m}\,  \textbf{t})  -\textbf{q}_j\right)
\langle \textbf{Q} |e^{-\beta\hat{K}}| {\textbf{R}'}\rangle
 \\ =
\langle \textbf{r}_1,\textbf{r}_2,\dots ,\textbf{r}_i  -i \tfrac{\hbar^2 \beta'}{m}\,  \textbf{t},\dots, \textbf{r}_j  +i \tfrac{\hbar^2 \beta'}{m}\,  \textbf{t},\dots, \textbf{r}_N|e^{-\beta\hat{K}}| 
\textbf{r}'_1,\textbf{r}'_2,\dots ,\textbf{r}'_i ,\dots, \textbf{r}'_j  ,\dots, \textbf{r}'_N
\rangle.
\end{multline}
The $N$-particle density matrix of non-interacting particles is the following \cite[Eq. (2.142)]{Feynman:1972:SMS}:
\begin{equation}
\label{eq:feynman}
\langle \textbf{r}_1,\textbf{r}_2,\dots ,\textbf{r}_i,\dots, \textbf{r}_j ,\dots, \textbf{r}_N|e^{-\beta\hat{K}}| 
\textbf{r}'_1,\textbf{r}'_2,\dots ,\textbf{r}'_i ,\dots, \textbf{r}'_j  ,\dots, \textbf{r}'_N
\rangle
 =
\left(\cfrac{m}{2\pi\hbar^2\beta}\right)^{\frac{3N}{2}}\exp\left(-\cfrac{m}{2\hbar^2\beta}\sum\limits_{n=1}^N(\textbf{r}_n-\textbf{r}'_n)^2\right).
\end{equation}
Substituting $\mathbf{r}_i\to \textbf{r}_i  -i \tfrac{\hbar^2 \beta'}{m}\,  \textbf{t}$,  $\mathbf{r}_j\to \textbf{r}_j  +i \tfrac{\hbar^2 \beta'}{m}\,  \textbf{t}$ in Eq.~\eqref{eq:feynman}, we obtain the matrix element \eqref{eq:dfhbej}:
\begin{multline}
\label{eq:hbdfiudj}
\langle \textbf{r}_1,\textbf{r}_2,\dots ,\textbf{r}_i  -i \tfrac{\hbar^2 \beta'}{m}\,  \textbf{t},\dots, \textbf{r}_j  +i \tfrac{\hbar^2 \beta'}{m}\,  \textbf{t},\dots, \textbf{r}_N|e^{-\beta\hat{K}}| 
\textbf{r}'_1,\textbf{r}'_2,\dots ,\textbf{r}'_i ,\dots, \textbf{r}'_j  ,\dots, \textbf{r}'_N
\rangle
\\=
\left(\cfrac{m}{2\pi\hbar^2\beta}\right)^{3N/2}\exp\left(-\cfrac{m}{2\hbar^2\beta}\sum\limits_{\substack{n=1\\n\neq i,j}}^N(\textbf{r}_n-\textbf{r}'_n)^2 -\cfrac{m}{2\hbar^2\beta}\left[(\textbf{r}_i  -i \tfrac{\hbar^2 \beta'}{m}\,  \textbf{t}-\textbf{r}'_i)^2 + ( \textbf{r}_j  +i \tfrac{\hbar^2 \beta'}{m}\,  \textbf{t}-\textbf{r}'_j)^2 \right]\right).
\end{multline}
Transforming the expression in $\left[\dots\right]$ in the exponent:
\begin{equation}
(\textbf{r}_i  -i \tfrac{\hbar^2 \beta'}{m}\,  \textbf{t}-\textbf{r}'_i)^2 + ( \textbf{r}_j  +i \tfrac{\hbar^2 \beta'}{m}\,  \textbf{t}-\textbf{r}'_j)^2
=
(\textbf{r}_i-\textbf{r}'_i)^2+(\textbf{r}_j-\textbf{r}'_j)^2 + 2i\cfrac{\hbar^2 \beta'}{m}\,\textbf{t} (\textbf{r}_j-\textbf{r}'_j) - 2i\cfrac{\hbar^2 \beta'}{m}\,\textbf{t} (\textbf{r}_i-\textbf{r}'_i)-2\cfrac{\hbar^4 \beta'^2}{m^2}\,t^2,
\end{equation}
we obtain \cite[Eq. (21)]{Kelbg:Annalen:1963}:
\begin{multline}
\label{eq:dfheij}
\langle \textbf{r}_1,\textbf{r}_2,\dots ,\textbf{r}_i  -i \tfrac{\hbar^2 \beta'}{m}\,  \textbf{t},\dots, \textbf{r}_j  +i \tfrac{\hbar^2 \beta'}{m}\,  \textbf{t},\dots, \textbf{r}_N|e^{-\beta\hat{K}}| 
\textbf{r}'_1,\textbf{r}'_2,\dots ,\textbf{r}'_i ,\dots, \textbf{r}'_j  ,\dots, \textbf{r}'_N
\rangle
\\=
\left(\cfrac{m}{2\pi\hbar^2\beta}\right)^{3N/2}\exp\left(-\cfrac{m}{2\hbar^2\beta}\sum\limits_{n=1}^N(\textbf{r}_n-\textbf{r}'_n)^2 -\cfrac{m}{2\hbar^2\beta}\left[
2i\tfrac{\hbar^2 \beta'}{m}\,\textbf{t} (\textbf{r}_j-\textbf{r}'_j) - 2i\tfrac{\hbar^2 \beta'}{m}\,\textbf{t} (\textbf{r}_i-\textbf{r}'_i)-2\tfrac{\hbar^4 \beta'^2}{m^2}\,t^2
\right]\right)
 \\ =
\left(\cfrac{m}{2\pi\hbar^2\beta}\right)^{3N/2}e^{-\tfrac{m}{2\hbar^2\beta}\sum\limits_{n=1}^N(\textbf{r}_n-\textbf{r}'_n)^2}
\exp\left(
\tfrac{\beta'}{\beta}i\,\textbf{t} (\textbf{r}_i-\textbf{r}'_i)
\right)
\exp\left(
-\tfrac{\beta'}{\beta}i\,\textbf{t} (\textbf{r}_j-\textbf{r}'_j)
\right)
\exp\left(
\tfrac{\hbar^2 \beta'^2}{\beta m}\,t^2
\right).
\end{multline}

To get the full density matrix, we need to calculate the first term in Eq. \eqref{eq:ghewjd}:
\begin{equation}
\langle \textbf{R} |e^{-\beta\hat{V}}e^{-\beta\hat{K}}| {\textbf{R}'}\rangle.
\end{equation}
To do it, we insert the coordinate variable $\mathbf{Q}$ and use Eqs. \eqref{eq:pot_matrx}, \eqref{eq:feynman}:
\begin{multline}
\langle \textbf{R} |e^{-\beta\hat{V}}\hat{1}e^{-\beta\hat{K}}| {\textbf{R}'}\rangle
=
\int d\textbf{Q} 
\langle \textbf{R} |e^{-\beta\hat{V}}| \textbf{Q}\rangle
\langle \textbf{Q}|
e^{-\beta\hat{K}}| {\textbf{R}'}\rangle=e^{-\beta U(\textbf{R})}\langle\textbf{R} |
e^{-\beta\hat{K}}| {\textbf{R}'}\rangle\\=
e^{-\beta U(\textbf{R})}\left(\cfrac{m}{2\pi\hbar^2\beta}\right)^{3N/2}
e^{-\tfrac{m}{2\hbar^2\beta}\sum\limits_{n=1}^N(\textbf{r}_n-\textbf{r}'_n)^2}.
\end{multline}
Finally, we write the full density matrix in the coordinate representation:
\begin{multline}
\label{eq:hejdk}
\langle \textbf{R} |\hat{\rho}(\beta)| {\textbf{R}'}\rangle
=
\left(\cfrac{m}{2\pi\hbar^2\beta}\right)^{3N/2}
e^{-\tfrac{m}{2\hbar^2\beta}\sum\limits_{n=1}^N(\textbf{r}_n-\textbf{r}'_n)^2}
e^{-\beta U(\textbf{R})}
\left\{1+\cfrac{1}{16\pi^3D}\sum_{i=1}^{N}\sum_{\substack{j = 1\\ j\neq i}}^{N}e_ie_j 
\int  v(\textbf{t}) e^{i\textbf{t}(\textbf{r}_i-\textbf{r}_j)}\right. \\ \times\left.
\int\limits_0^{\beta}\beta_1 \cfrac{d }{d  \beta_1}
\left(
\exp\left(
\tfrac{\beta'}{\beta}i\,\textbf{t} (\textbf{r}_i-\textbf{r}'_i)
\right)
\exp\left(
-\tfrac{\beta'}{\beta}i\,\textbf{t} (\textbf{r}_j-\textbf{r}'_j)
\right)
e^{\tfrac{\hbar^2(\beta_1-\beta)}{m}\,t^2}
\exp\left(
\tfrac{\hbar^2 \beta'^2}{\beta m}\,t^2
\right)
\right)d\beta_1d\textbf{t}
\right\}.
\end{multline}

\subsection{Transformation of density matrix Eq. \eqref{eq:hejdk}}
Next, we change the variables in the integral:
\begin{equation}
\alpha = \beta_1/\beta, \quad \beta d\alpha = d\beta_1.
\end{equation}
Then the integral over $\beta_1$ in Eq. \eqref{eq:hejdk} turns into the following form:
\begin{multline}
\int\limits_0^{\beta}\beta_1 \cfrac{d }{d  \beta_1}
\left(
\exp\left(
\tfrac{\beta'}{\beta}i\,\textbf{t} (\textbf{r}_i-\textbf{r}'_i)
\right)
\exp\left(
-\tfrac{\beta'}{\beta}i\,\textbf{t} (\textbf{r}_j-\textbf{r}'_j)
\right)
e^{\tfrac{\hbar^2(\beta_1-\beta)}{m}\,t^2}
\exp\left(
\tfrac{\hbar^2 \beta'^2}{\beta m}\,t^2
\right)
\right)d\beta_1
\\
=
\beta\int\limits_0^{1}\alpha \cfrac{d }{d  \alpha }
\left(
\exp\left(
(\alpha-1)i\,\textbf{t} (\textbf{r}_i-\textbf{r}'_i)
\right)
\exp\left(
-(\alpha-1)i\,\textbf{t} (\textbf{r}_j-\textbf{r}'_j)
\right)
e^{\tfrac{\hbar^2(\alpha-1)\beta}{m}\,t^2}
\exp\left(
\tfrac{\hbar^2 (\alpha-1)^2\beta}{ m}\,t^2
\right)
\right)d\alpha.
\end{multline}
Making the following transformations:
\begin{multline}
\exp\left(
(\alpha-1)i\,\textbf{t} (\textbf{r}_i-\textbf{r}'_i)
\right)
\exp\left(
-(\alpha-1)i\,\textbf{t} (\textbf{r}_j-\textbf{r}'_j)
\right)
e^{\tfrac{\hbar^2(\alpha-1)\beta}{m}\,t^2}
\exp\left(
\tfrac{\hbar^2 (\alpha-1)^2\beta}{ m}\,t^2
\right)
 \\ =
\exp\bigl[
\alpha i\,\textbf{t} (\textbf{r}_i-\textbf{r}'_i) - i\,\textbf{t} (\textbf{r}_i-\textbf{r}'_i)
\bigr]
\exp\bigl[
-\alpha i\,\textbf{t} (\textbf{r}_j-\textbf{r}'_j) + i\,\textbf{t} (\textbf{r}_j-\textbf{r}'_j)
\bigr]
\exp\left(
\tfrac{(\alpha-1)\alpha\hbar^2 \beta}{ m}\,t^2
\right)
 \\ =
\exp\left[
\alpha i\,\textbf{t} (\textbf{r}_i-\textbf{r}'_i) -\alpha i\,\textbf{t} (\textbf{r}_j-\textbf{r}'_j)
\right]
\exp\left(-
\tfrac{(1-\alpha)\alpha\hbar^2 \beta}{ m}\,t^2
\right)
\exp\left(
- i\,\textbf{t} (\textbf{r}_i-\textbf{r}'_i)+ i\,\textbf{t} (\textbf{r}_j-\textbf{r}'_j)
\right),
\end{multline}
we can rewrite the integral over $\textbf{t}$ and $\alpha$ in Eq. \eqref{eq:hejdk}:
\begin{multline}
\int  v(\textbf{t}) e^{i\textbf{t}(\textbf{r}_i-\textbf{r}_j)}
\int\limits_0^{\beta}\beta_1 \cfrac{d }{d  \beta_1}
\left(
\exp\left(
\tfrac{\beta'}{\beta}i\,\textbf{t} (\textbf{r}_i-\textbf{r}'_i)
\right)
\exp\left(
-\tfrac{\beta'}{\beta}i\,\textbf{t} (\textbf{r}_j-\textbf{r}'_j)
\right)
\exp\left(
\tfrac{\hbar^2 \beta'^2}{\beta m}\,t^2
\right)
\right)d\beta_1d\textbf{t}
 \\ =
\int  v(\textbf{t}) e^{i\textbf{t}(\textbf{r}'_i-\textbf{r}'_j)}
\beta\int\limits_0^{1}\alpha \cfrac{d }{d  \alpha }\left(
e^{
\alpha i\,\textbf{t} (\textbf{r}_i-\textbf{r}'_i) -\alpha i\,\textbf{t} (\textbf{r}_j-\textbf{r}'_j)
}
e^{-
\tfrac{(1-\alpha)\alpha\hbar^2 \beta}{ m}\,t^2 
}
\right)d\alpha d\textbf{t}
\end{multline}
to obtain \cite[Eq. (22)]{Kelbg:Annalen:1963}:
\begin{multline}
\label{eq:fshbaiuh}
\langle \textbf{R} |\hat{\rho}(\beta)| {\textbf{R}'}\rangle
=
\left(\cfrac{m}{2\pi\hbar^2\beta}\right)^{3N/2}
e^{-\tfrac{m}{2\hbar^2\beta}\sum\limits_{n=1}^N(\textbf{r}_n-\textbf{r}'_n)^2}
e^{-\beta U(\textbf{R})}
\left\{1+\cfrac{\beta}{16\pi^3D}\sum_{i=1}^{N}\sum_{\substack{j = 1\\ j\neq i}}^{N}e_ie_j 
\int  v(\textbf{t}) e^{i\textbf{t}(\textbf{r}'_i-\textbf{r}'_j)} \right. \\ \times\left.
\int\limits_0^{1}\alpha \cfrac{d }{d  \alpha }\left(
e^{
i\alpha \textbf{t} (\textbf{r}_{i}-\textbf{r}_{j}-\textbf{r}'_{i}+\textbf{r}'_{j})
}
e^{-\alpha(1-\alpha)
\tfrac{\hbar^2 \beta}{ m}\,t^2 
}
\right)d\alpha d\textbf{t}
\right\}.
\end{multline}
We introduce the notation:
\begin{equation}
\textbf{r}_{ij} = \textbf{r}_i-\textbf{r}_j, \quad  \textbf{r}'_{ij} = \textbf{r}'_i-\textbf{r}'_j.
\end{equation}
Now we integrate the term in Eq. \eqref{eq:fshbaiuh} over $\alpha$ by parts:
\begin{multline}
\int\limits_0^{1}\alpha \cfrac{d }{d  \alpha }\left(
e^{
i\alpha \textbf{t} (\textbf{r}_{ij}-\textbf{r}'_{ij})
}
e^{-\alpha(1-\alpha)
\tfrac{\hbar^2 \beta}{ m}\,t^2 
}
\right)d\alpha = 
\left. \alpha\left(
e^{
i\alpha \textbf{t} (\textbf{r}_{ij}-\textbf{r}'_{ij})
}
e^{-\alpha(1-\alpha)
\tfrac{\hbar^2 \beta}{ m}\,t^2 
}
\right)\right|_0^1 - \int\limits_0^{1}e^{
i\alpha \textbf{t} (\textbf{r}_{ij}-\textbf{r}'_{ij})
}
e^{-\alpha(1-\alpha)
\tfrac{\hbar^2 \beta}{ m}\,t^2 
}
d\alpha
 \\ =
e^{
i\textbf{t} (\textbf{r}_{ij}-\textbf{r}'_{ij})
}
- \int\limits_0^{1}e^{
i\alpha \textbf{t} (\textbf{r}_{ij}-\textbf{r}'_{ij})
}
e^{-\alpha(1-\alpha)
\tfrac{\hbar^2 \beta}{ m}\,t^2 
}
d\alpha.
\end{multline}
Thus the additional term in Eq. \eqref{eq:fshbaiuh} becomes the following:
\begin{multline}
\label{eq:fdhies}
\cfrac{\beta}{16\pi^3D}\sum_{i=1}^{N}\sum_{\substack{j = 1\\ j\neq i}}^{N}e_ie_j 
\int  v(\textbf{t}) e^{i\textbf{t}\textbf{r}'_{ij}} 
\int\limits_0^{1}\alpha \cfrac{d }{d  \alpha }\left(
e^{
i\alpha \textbf{t} (\textbf{r}_{ij}-\textbf{r}'_{ij})
}
e^{-\alpha(1-\alpha)
\tfrac{\hbar^2 \beta}{ m}\,t^2 
}
\right)d\alpha d\textbf{t}  \\ =
\cfrac{\beta}{16\pi^3D}\sum_{i=1}^{N}\sum_{\substack{j = 1\\ j\neq i}}^{N}e_ie_j 
\int  v(\textbf{t}) e^{i\textbf{t}\textbf{r}'_{ij}} 
e^{
i\textbf{t} (\textbf{r}_{ij}-\textbf{r}'_{ij})
} d\textbf{t} \\
{} - \cfrac{\beta}{16\pi^3D}\sum_{i=1}^{N}\sum_{\substack{j = 1\\ j\neq i}}^{N}e_ie_j 
\int  v(\textbf{t}) e^{i\textbf{t}\textbf{r}'_{ij}} 
\int\limits_0^{1}e^{
i\alpha \textbf{t} (\textbf{r}_{ij}-\textbf{r}'_{ij})
}
e^{-\alpha(1-\alpha)
\tfrac{\hbar^2 \beta}{ m}\,t^2 
}
d\alpha d\textbf{t}.
\end{multline}
Due to the first term in Eq. \eqref{eq:fdhies}, the potential energy appears:
\begin{equation}
\cfrac{\beta}{16\pi^3D}\sum_{i=1}^{N}\sum_{\substack{j = 1\\ j\neq i}}^{N}e_ie_j 
\int  v(\textbf{t}) e^{i\textbf{t}\textbf{r}'_{ij}} 
e^{
i\textbf{t} (\textbf{r}_{ij}-\textbf{r}'_{ij})
} d\textbf{t}
= 
\cfrac{\beta}{16\pi^3D}\sum_{i=1}^{N}\sum_{\substack{j = 1\\ j\neq i}}^{N}e_ie_j 
\int  v(\textbf{t}) e^{i\textbf{t}\textbf{r}_{ij}} 
 d\textbf{t} = \beta U(\textbf{R}).
\end{equation}

Let us consider the second term in Eq. \eqref{eq:fdhies} 
and rewrite the integral:
\begin{equation}
\label{eq:rehgewnjs}
\int  v(\textbf{t}) e^{i\textbf{t}\textbf{r}'_{ij}} 
\int\limits_0^{1}e^{
i\alpha \textbf{t} (\textbf{r}_{ij}-\textbf{r}'_{ij})
}
e^{-\alpha(1-\alpha)
\tfrac{\hbar^2 \beta}{ m}\,t^2 
}
d\alpha d\textbf{t}
=
\int\limits_0^{1}d\alpha
\int  v(\textbf{t})
e^{
i \textbf{t}[\alpha\textbf{r}_{ij}+(1-\alpha)\textbf{r}'_{ij}] 
}
e^{-\alpha(1-\alpha)
\tfrac{\hbar^2 \beta}{ m}\,t^2 
}
 d\textbf{t}.
\end{equation}
Introducing one more notation:
\begin{equation}
\textbf{d}_{ij}(\alpha) = \alpha\textbf{r}_{ij}+(1-\alpha)\textbf{r}'_{ij}, \quad d_{ij}(\alpha) = |\alpha\textbf{r}_{ij}+(1-\alpha)\textbf{r}'_{ij}|,
\end{equation}
we write the integral in Eq. \eqref{eq:rehgewnjs} over $\textbf{t}$ as follows:
\begin{equation}
\int  v(\textbf{t})
e^{
i \textbf{t}[\alpha\textbf{r}_{ij}+(1-\alpha)\textbf{r}'_{ij}] 
}
e^{-\alpha(1-\alpha)
\tfrac{\hbar^2 \beta}{ m}\,t^2 
}
 d\textbf{t}=
4\pi\int \cfrac{1}{t^2}\,
e^{
i \textbf{t}\textbf{d}_{ij}(\alpha) 
}
e^{-\alpha(1-\alpha)
\tfrac{\hbar^2 \beta}{ m}\,t^2 
}
 d\textbf{t}.
\end{equation}

Thus, the density matrix \eqref{eq:fshbaiuh} has the form:
\begin{equation}
\label{eq:fdhshj}
\langle \textbf{R} |\hat{\rho}(\beta)| {\textbf{R}'}\rangle
=
\left(\cfrac{m}{2\pi\hbar^2\beta}\right)^{3N/2}
e^{-\tfrac{m}{2\hbar^2\beta}\sum\limits_{n=1}^N(\textbf{r}_n-\textbf{r}'_n)^2}
e^{-\beta U(\textbf{R})}
\left\{1+
\beta U(\textbf{R})
-
\cfrac{\beta}{2}\sum_{i=1}^{N}\sum_{\substack{j = 1\\ j\neq i}}^{N}\cfrac{e_ie_j}{D} \Phi(\textbf{r}_{ij},\textbf{r}'_{ij}; \beta)
\right\},
\end{equation}
where
\begin{equation}
\label{eq:phidef}
\Phi(\textbf{r}_{ij},\textbf{r}'_{ij}; \beta) = 
\cfrac{1}{8\pi^3}\int\limits_0^{1}d\alpha
\int  v(\textbf{t})
e^{
i \textbf{t}\textbf{d}_{ij}(\alpha) 
}
e^{-\alpha(1-\alpha)
\lambda^2t^2 
}
 d\textbf{t}, \quad \lambda^2 = \lambda^2(\beta) =\cfrac{\hbar^2\beta}{m}.
\end{equation}
One can see that different interaction potentials $v(\textbf{t})$ produce different functions $\Phi(\textbf{r}_{ij},\textbf{r}'_{ij};\beta)$. Due to it, we further call Eq. \eqref{eq:phidef} \emph{the Kelbg functional}. The function $\Phi(\textbf{r}_{ij},\textbf{r}_{ij};\beta)$ is \emph{the diagonal Kelbg functional}.

Let us consider the function $e^x$. If $x\ll 1$ we can write $e^x\approx 1 + x$. We have obtained Eq. \eqref{eq:fdhshj} in the first order of $\hat{V}$. Thus the following quantity should be much less than $1$:
\begin{equation}
\left[\beta U(\textbf{R})
-
\cfrac{\beta}{2}\sum_{i=1}^{N}\sum_{\substack{j = 1\\ j\neq i}}^{N}\cfrac{e_ie_j}{D} \Phi(\textbf{r}_{ij},\textbf{r}'_{ij};\beta) \right]\ll 1.
\end{equation}
This is the requirement for the perturbation theory to be applicable. So we can formally use the equivalence of $1 + x$ and $e^x$ for a small $x$:
\begin{equation}
1+
\beta U(\textbf{R})
-
\cfrac{\beta}{2}\sum_{i=1}^{N}\sum_{\substack{j = 1\\ j\neq i}}^{N}\cfrac{e_ie_j}{D} \Phi(\textbf{r}_{ij},\textbf{r}'_{ij}; \beta) \approx e^{\beta U(\textbf{R})}
\exp\left\{
-
\frac{\beta}{2}\sum_{i=1}^{N}\sum_{\substack{j = 1\\ j\neq i}}^{N}\frac{e_ie_j}{D} \Phi(\textbf{r}_{ij},\textbf{r}'_{ij}; \beta)\right\}.
\end{equation}
Substituting it in Eq. \eqref{eq:fdhshj}, we obtain:
\begin{multline}
\langle \textbf{R} |\hat{\rho}(\beta)| {\textbf{R}'}\rangle
=
\left(\cfrac{m}{2\pi\hbar^2\beta}\right)^{3N/2}
e^{-\tfrac{m}{2\hbar^2\beta}\sum\limits_{n=1}^N(\textbf{r}_n-\textbf{r}'_n)^2}
e^{-\beta U(\textbf{R})}
e^{\beta U(\textbf{R})}\exp\left\{
-
\frac{\beta}{2}\sum_{i=1}^{N}\sum_{\substack{j = 1\\ j\neq i}}^{N}\frac{e_ie_j}{D} \Phi(\textbf{r}_{ij},\textbf{r}'_{ij}; \beta)\right\} \\ =
\left(\cfrac{m}{2\pi\hbar^2\beta}\right)^{3N/2}
e^{-\tfrac{m}{2\hbar^2\beta}\sum\limits_{n=1}^N(\textbf{r}_n-\textbf{r}'_n)^2}\exp\left\{
-
\frac{\beta}{2}\sum_{i=1}^{N}\sum_{\substack{j = 1\\ j\neq i}}^{N}\frac{e_ie_j}{D} \Phi(\textbf{r}_{ij},\textbf{r}'_{ij}; \beta)\right\}.
\end{multline}

\section{Derivation of Kelbg pseudopotential}
\subsection{Diagonal Kelbg pseudopotential}
Let us consider the diagonal matrix elements:
\begin{equation}
\langle \textbf{R} |\hat{\rho}(\beta)| {\textbf{R}}\rangle
=
\left(\cfrac{m}{2\pi\hbar^2\beta}\right)^{3N/2}\exp\left\{
-
\frac{\beta}{2}\sum_{i=1}^{N}\sum_{\substack{j = 1\\ j\neq i}}^{N}\frac{e_ie_j}{D} \Phi(\textbf{r}_{ij},\textbf{r}_{ij};\beta)\right\}.
\end{equation}
Actually, it is \cite[Eq. (23)]{Kelbg:Annalen:1963}.
To compute $\Phi(\textbf{r}_{ij},\textbf{r}_{ij};\beta)$, we integrate Eq. \eqref{eq:phidef} over $\alpha$:
\begin{equation}
\label{eq:ghewhikfj}
\Phi(\textbf{r}_{ij},\textbf{r}_{ij};\beta) = 
\cfrac{1}{8\pi^3}
\int  d\textbf{t}  v(\textbf{t})
e^{
i \textbf{t}\textbf{r}_{ij}(\alpha) 
}
\int\limits_0^{1}d\alpha
e^{-\alpha(1-\alpha)
\lambda^2t^2 
},
\end{equation}
\begin{equation}
\int\limits_0^1e^{-\alpha(1-\alpha)
\lambda^2 t^2 
}d\alpha
=
e^{-\lambda^2t^2/4}\sqrt{\pi}\,\cfrac{\mathrm{erfi}(\lambda t / 2)}{\lambda t},
\end{equation}
where $\mathrm{erfi}(x)= -i \mathrm{erf}(ix)$ and $\mathrm{erf}(x)$ is the error function. 

Now substituting the Fourier component of the Coulomb potential:
\begin{equation}
\label{eq:coulumbFourier}
v(\textbf{t}) = \cfrac{4\pi}{t^2}
\end{equation}
in Eq. \eqref{eq:ghewhikfj}, we integrate over $\textbf{t}$ in the spherical coordinates:
\begin{multline}
\label{eq:kelgbpot}
\Phi(\textbf{r}_{ij},\textbf{r}_{ij};\beta) = 
\cfrac{4\pi}{8\pi^3}
\int  d\textbf{t}  \cfrac{1}{t^2}
e^{
i \textbf{t}\textbf{r}_{ij}(\alpha) 
}
e^{-\lambda^2t^2/4}\sqrt{\pi}\,\cfrac{\mathrm{erfi}(\lambda t / 2)}{\lambda t} 
 \\=
\cfrac{2}{\sqrt{\pi}} \int\limits_0^{\infty}e^{-\lambda^2t^2/4}\cfrac{\sin(t r_{ij})}{t r_{ij}}\,\cfrac{\mathrm{erfi}(\lambda t / 2)}{\lambda t} = \cfrac{1}{r_{ij}}\left( 1-e^{-r_{ij}^2/\lambda^2}+\sqrt{\pi} r_{ij}/\lambda [1-\mathrm{erf}(r_{ij}/\lambda)]\right).
\end{multline}
The expression \eqref{eq:kelgbpot} is often called the ``Kelbg potential'' or ``Kelbg pseudopotential''.

\subsection{Non-diagonal Kelbg pseudopotential}
Next, we consider again Eq. \eqref{eq:phidef} to write it in a more compact form. We first calculate the integral over $\textbf{t}$ in the spherical coordinates, using Eq. \eqref{eq:coulumbFourier}:
\begin{equation}
\label{eq:hfsdjskj}
\int  v(\textbf{t})
e^{
i \textbf{t}\textbf{d}_{ij}(\alpha) 
}
e^{-\alpha(1-\alpha)
\lambda^2t^2 
}
 d\textbf{t} = 16\pi^2 \int\limits_0^{\infty} \cfrac{\sin(t d_{ij}(\alpha) )}{t d_{ij}(\alpha)} e^{-\alpha(1-\alpha)
\lambda^2t^2 
}d t
=
\cfrac{8\pi^3}{d_{ij}(\alpha)} 
\mathrm{erf}\left(\cfrac{d_{ij}(\alpha)/\lambda}{2\sqrt{\alpha(1-\alpha)}}\right).
\end{equation}
Substituting Eq. \eqref{eq:hfsdjskj} in Eq. \eqref{eq:phidef}, we obtain the non-diagonal Kelbg pseudopotential:
\begin{equation}
\Phi(\textbf{r}_{ij},\textbf{r}'_{ij}; \beta) = 
\int\limits_0^{1}
\cfrac{d\alpha}{d_{ij}(\alpha)} 
\mathrm{erf}\left(\cfrac{d_{ij}(\alpha)/\lambda}{2\sqrt{\alpha(1-\alpha)}}\right).
\end{equation}
If the particles have different masses, we must replace the mass in $\lambda$ by the reduced mass:
\begin{equation}
\lambda^2 = \cfrac{\hbar^2\beta}{m} \Rightarrow \cfrac{\hbar^2\beta}{2\mu_{ij}} = \lambda^2_{ij}, \quad \mu^{-1}_{ij} = m^{-1}_i + m^{-1}_j.
\end{equation}

\section{Conclusion}
In this report we have presented an extensive derivation of the density matrix for a system of Coulomb particles in the high--temperature limit. We have followed the original work by Kelbg \cite{Kelbg:Annalen:1963} but restored many details skipped in the original paper. We hope that our efforts will be useful for researchers in the field of quantum statistical physics.



\appendix

\section{Truncation of series Eq. \eqref{eq:linearGdiff}}
\label{sec:app1}
In this section, we calculate the commutator $[\hat{V},[\hat{V},[\hat{V},\hat{K}]]]$. We work here in the coordinate representation. The potential energy $U(\textbf{R})$ is rewritten in the terms of a function $v(\textbf{r}_i)$:
\begin{equation}
U(\textbf{R}) = \cfrac{1}{2}\sum_{i=1}^N\sum_{\substack{j= 1\\ j\neq i}}^Nu_{ij}(\textbf{r}_i - \textbf{r}_j) \equiv  \sum_{i=1}^N v(\textbf{r}_i).
\end{equation}
Let us calculate $[\hat{V}, \hat{K}]$:
\begin{equation}
[\hat{V}, \hat{K}]\Psi(\textbf{R}) = \cfrac{\hbar^2}{2m}\left[\sum_{i = 1}^N\nabla_i^2, \sum_{j = 1}^Nv(\textbf{r}_j)\right]\Psi(\textbf{R}) = 
\left(\cfrac{\hbar^2}{2m}\right) \sum_{i = 1}^N\sum_{j = 1}^N\left[\nabla_i^2, v(\textbf{r}_j)\right]\Psi(\textbf{R}).
\end{equation}
Consider each term individually:
\begin{equation}
\left[\nabla_i^2, v(\textbf{r}_j)\right]\Psi(\textbf{R}) = \nabla_i^2v(\textbf{r}_j)\Psi(\textbf{R}) - v(\textbf{r}_j)\nabla_i^2\Psi(\textbf{R}).
\end{equation}
The differentiation produces three components:
\begin{equation}
\nabla_i^2v(\textbf{r}_j)\Psi(\textbf{R})
=
\Psi(\textbf{R}) (\nabla_i^2v(\textbf{r}_i))\delta_{ij}+
2 (\nabla_iv(\textbf{r}_i) )(\nabla_i \Psi(\textbf{R}))\delta_{ij} + 
v(\textbf{r}_j)\nabla_i^2\Psi(\textbf{R}),
\end{equation}
where $\delta_{ij}$ is the Kronecker delta.
Thus, we obtain:
\begin{equation}
[\hat{V}, \hat{K}] = \cfrac{\hbar^2}{2m}\sum_{i = 1}^N\left\{(\nabla_i^2v(\textbf{r}_i)) + 2 (\nabla_iv(\textbf{r}_i) )\nabla_i\right\}.
\end{equation}

Let us now calculate $[\hat{V},[\hat{V}, \hat{K}]]$:
\begin{multline}
[\hat{V},[\hat{V}, \hat{K}]] = \cfrac{\hbar^2}{2m}\left[ \sum_{j = 1}^Nv(\textbf{r}_j), \sum_{i = 1}^N\left\{(\nabla_i^2v(\textbf{r}_i)) + 2 (\nabla_iv(\textbf{r}_i) )\nabla_i\right\}\right] = 
\cfrac{\hbar^2}{m}\left[ \sum_{j = 1}^Nv(\textbf{r}_j), \sum_{i = 1}^N (\nabla_iv(\textbf{r}_i) )\nabla_i\right]\\
=
\cfrac{\hbar^2}{m}\sum_{j = 1}^N\sum_{i = 1}^N\left[ v(\textbf{r}_j),  (\nabla_iv(\textbf{r}_i) )\nabla_i\right].
\end{multline}
Again, we consider each term individually:
\begin{equation}
\left[ v(\textbf{r}_j),  (\nabla_iv(\textbf{r}_i) )\nabla_i\right]\Psi(\textbf{R}) = 
v(\textbf{r}_j) (\nabla_iv(\textbf{r}_i) )\nabla_i\Psi(\textbf{R})
-
(\nabla_iv(\textbf{r}_i) )\nabla_iv(\textbf{r}_j) \Psi(\textbf{R}).
\end{equation}
The differentiation produces two components:
\begin{equation}
\nabla_iv(\textbf{r}_j) \Psi(\textbf{R})
=
\Psi(\textbf{R}) (\nabla_iv(\textbf{r}_i)) \delta_{ij} + v(\textbf{r}_j) (\nabla_i\Psi(\textbf{R})).
\end{equation}
Thus, we obtain:
\begin{equation}
[\hat{V},[\hat{V}, \hat{K}]] =  -\cfrac{\hbar^2}{m}\sum_{i = 1}^N (\nabla_iv(\textbf{r}_i) )^2.
\end{equation}
Note that $[\hat{V},[\hat{V}, \hat{K}]]$ is the function only of a coordinate $\textbf{R}$.
Thus, it commutates with the potential energy:
\begin{equation}
[\hat{V},[\hat{V},[\hat{V},\hat{K}]]] = -\cfrac{\hbar^2}{m}\sum_{j = 1}^N\sum_{i = 1}^N\left[v(\textbf{r}_j), (\nabla_iv(\textbf{r}_i) )^2\right] = 0.
\end{equation}


\end{document}